\documentclass[%
twocolumn,
superscriptaddress,
amsmath,amssymb,
aps,
prx
]{revtex4-2}

\usepackage{stmaryrd} %
\usepackage{graphicx} 
\usepackage{float}
\usepackage[T1]{fontenc}
\usepackage{braket}
\usepackage{xcolor}
\usepackage[normalem]{ulem} 
\usepackage{mathtools}
\usepackage{multirow}
\usepackage{url}
\usepackage{hyperref}
\usepackage{amsmath,amsthm,amssymb}
\usepackage{bm}

\AtBeginDocument{\renewcommand{\selectlanguage}[1]{}}
\newtheorem{remark}{Remark}

\begin{document}
\title{
Strictly Local Tile-Code Architectures on Two-Dimensional Planar Lattices
}
\author{Yoonjin Bae}
\affiliation{Department of Physics and Astronomy, Seoul National University, Seoul 08826, South Korea}

\author{Chae-Yeun Park}
\affiliation{School of Integrated Technology, Yonsei University, Seoul 03722, South Korea}
\affiliation{Department of Quantum Information, Yonsei University, Seoul 03722, South Korea}
\affiliation{BK21 Graduate Program in Intelligent Semiconductor Technology, Yonsei University, Seoul 03722, South Korea}

\date{\today}

\begin{abstract}
Tile codes are a family of planar quantum low-density parity-check (qLDPC) codes with weight-6 stabilizers and open boundary conditions, offering an encoding efficiency $kd^2/n$ of up to four times that of the surface code.
In this work, we develop an exhaustive search algorithm for finding \texttt{SWAP}-based routing schemes that implement syndrome extraction for four tile-code families using only nearest-neighbor interactions on a two-dimensional square lattice, matching the connectivity of the surface code.
Using explicitly constructed routed syndrome-extraction circuits decoded with BP+OSD, we estimate the circuit-level thresholds of these code families.
For the SI1000 noise model, the threshold without such a connectivity constraint is obtained in a range $0.23\%$--$0.31\%$, while it decreases to $0.11\%$--$0.13\%$ with routing, representing a reduction factor of around two to three.
Despite this threshold penalty, our resource-footprint analysis shows that routed tile codes require fewer physical qubits per logical qubit than the surface code at sufficiently low physical error rates: Under the SI1000 noise model, we find a crossover near $p^*\approx 0.08\%$, below which routed tile codes become more qubit-efficient, with an advantage that grows monotonically as the physical error rate decreases.
\end{abstract}
\maketitle

\section{Introduction}

Quantum computers are believed to solve certain classically intractable problems more efficiently than classical computers.
Shor's algorithm is the most prominent example, solving integer factorization in polynomial time~\cite{shor1994algorithms}.
However, realizing such algorithms requires quantum circuits with a large number of gates, making them highly sensitive to noise in experimentally available qubits. 
Fault-tolerant quantum computation addresses this challenge by encoding logical qubits into many physical qubits and repeatedly extracting error syndromes to detect and correct faults without directly measuring the encoded quantum information~\cite{shor1996fault,aharonov1997fault}.
In this scheme, the logical error rate can be arbitrarily suppressed by increasing the code distance, provided that the physical error rate is below the threshold.

For decades, the toric code, surface code, and their variants have been leading candidates for fault-tolerant quantum computation because they use geometrically local, low-weight stabilizer checks on a two-dimensional architecture and exhibit high thresholds under realistic noise models~\cite{dennis2002topological, fowler2012surface,terhal2015quantum}.
These properties make surface-code-based architectures particularly attractive for near-term hardware. 
However, surface codes encode only a small number of logical qubits per patch, so achieving large code distances requires a substantial number of physical qubits.
This overhead has motivated the search for quantum error-correcting codes that retain favorable syndrome-extraction properties while improving encoding efficiency.

Quantum low-density parity-check (qLDPC) codes have recently emerged as a promising candidate for fault-tolerant quantum memory with low overhead and high thresholds~\cite{Breuckmann2021, Panteleev2021, Tremblay2022}.
In particular, \emph{Bivariate Bicycle (BB) codes} introduced by \citet{bravyi24high} demonstrate a compelling combination of high encoding efficiency and a logical error pseudo-threshold of approximately $0.8\%$, achieved using a low-depth, degree-$6$ syndrome-extraction circuit.
These results position BB codes as one of the most hardware-efficient LDPC-based quantum memories known to date.

Despite these advantages, BB codes rely critically on (1) periodic boundary conditions and (2) non-local qubit connectivity.
In contrast, the currently available superconducting qubit architecture is inherently two-dimensional, and long-range interactions are costly or unavailable.
This gap between theoretical performance and hardware realization motivates the search for qLDPC codes that retain the favorable properties of BB codes while relaxing their architectural constraints.

Several recent works tried to close this gap in two complementary directions.
On the one hand, hardware-efficient implementations of syndrome-measurement circuits for BB codes have been proposed~\cite{shaw2025lowering,zhao2025simple,mathews2026placing,berthusen2025toward,liu2026efficient}.
These suggestions develop routing schemes under certain hardware constraints, but most current suggestions still assume a multi-layer architecture~\cite{shaw2025lowering,mathews2026placing,berthusen2025toward} or long-range connectivity~\cite{zhao2025simple}.
On the other hand, new qLDPC code families with lower locality constraints are also developed.
For example, \citet{geher2025directional} introduced directional codes, where the code is more inversely constructed from the hardware constraint---the nearest-neighbor interaction.
However, directional codes assumed a periodic 2D lattice, opening the question of codes on open boundaries. 
\citet{liang_planar_2025} and \citet{steffan_tile_2025} proposed families of qLDPC codes that are implementable in a planar lattice with open boundaries while largely preserving the algebraic structure of BB codes. 
Still, the stabilizers of these codes are not strictly local on a two-dimensional planar lattice, and a naive implementation of syndrome-extraction circuits requires mid-range connectivity between ancilla and data qubits.

In this work, we develop a routing scheme for four tile code families on an architecture with strictly local connectivity and compare circuit-level thresholds with and without this constraint.
The stabilizers of these codes are local in small windows of sizes $3\times 3$ and $4 \times 3$ on a 2D square lattice, and the encoding efficiency $kd^2/n$ ranges from $1.78$ to $4.00$.
We refer to all these families collectively as tile codes throughout this paper.
We consider both standard depolarizing and SI1000 noise models, as well as a routing scheme and a syndrome measurement circuit implementable only using nearest-neighbor gates on a 2D lattice.
Most significantly, using the BP+OSD decoder, we estimate the threshold under the SI1000 noise model to be $0.23\%-0.31\%$ when two-qubit gates can be applied without the locality constraints, and $0.11\%-0.13\%$ with routing, where two-qubit gates only between nearest-neighbor qubits on a 2D square lattice are allowed.

We also analyze the resource footprint of routed tile codes compared with the rotated surface code.
Despite the lower thresholds of routed tile codes, we find explicit instances in the $[[288,8,12]]$ family that are more resource-efficient than the rotated surface code when $p\lesssim 0.08\%$ under the SI1000 noise model.
Our results suggest that planar qLDPC codes, albeit their long-range connectivity, can be a practical for fault-tolerant quantum computer with proper routing scheme.

\begin{figure}
    \centering
    \includegraphics[width=1\linewidth]{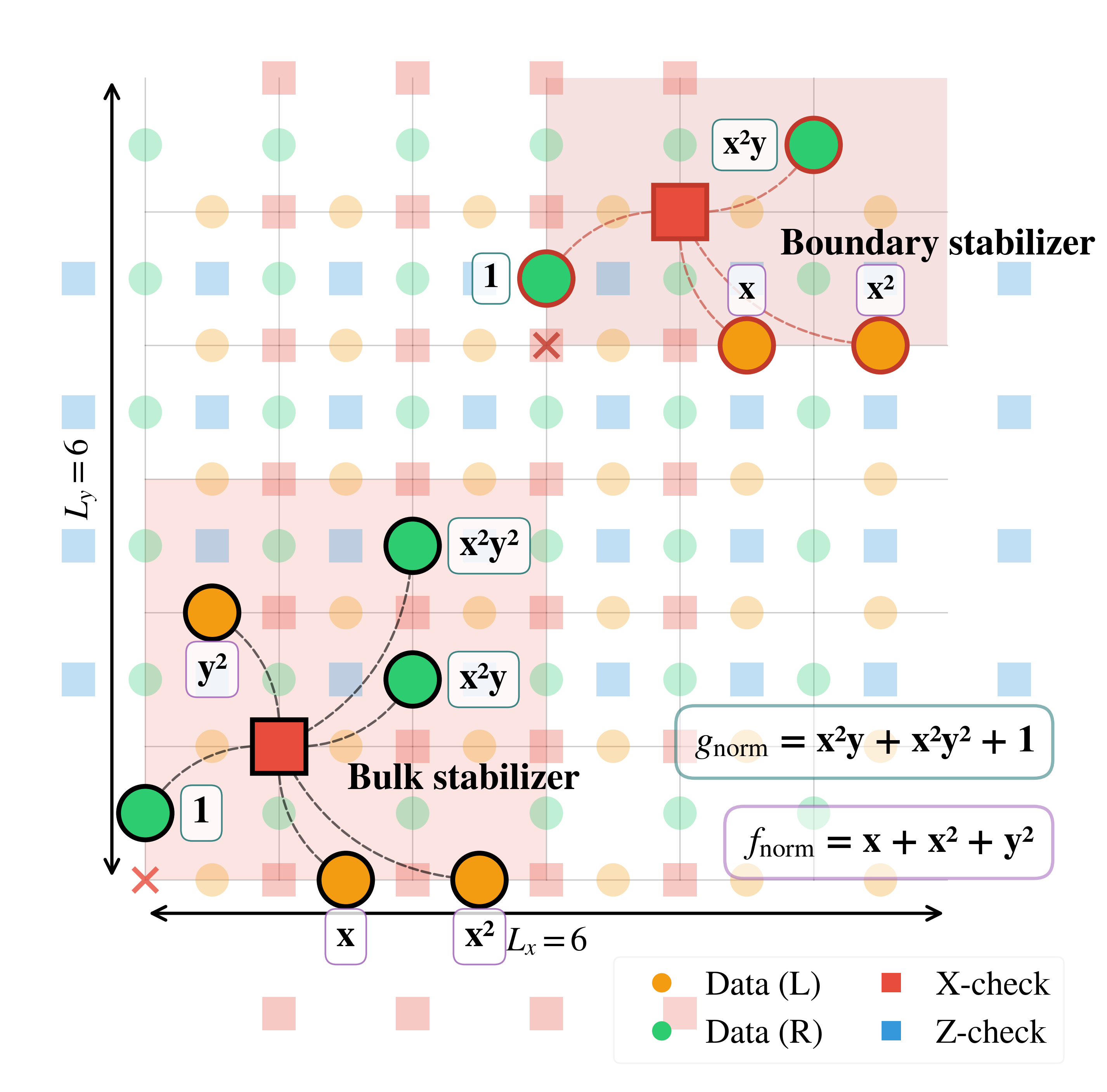}
    \caption{Illustration of the [[288, 8, 12]] tilecode family with $L_x = L_y = 6$ and distance $d=4$. The lattice structure shows data qubits (circles) positioned on edges: Data \texttt{L} qubits (orange) on horizontal edges and Data \texttt{R} qubits (green) on vertical edges. Check operators (squares) are positioned at vertices and faces: $X$-check operators (red) and $Z$-check operators (blue). Two representative $X$-type stabilizers are highlighted: a bulk stabilizer with weight 6 (black border) and a boundary stabilizer with weight 4 (red border), each connected to its support qubits via dashed curves. The semi-transparent shaded regions indicate the $3 \times 3$ lattice support areas, with \textsf{X} markers denoting the origin points corresponding to the constant term in the polynomials.}
    \label{fig:tilecode_structure}
\end{figure}

\section{Preliminaries}

\subsection{Tile Code Families}
\textbf{Bulk and boundary stabilizers.} The stabilizers of each tile code can be classified into \emph{bulk stabilizers}, whose full stabilizer support lies on qubit edges in the interior of the lattice, and \emph{boundary stabilizers}, which arise from truncated bulk stabilizers near open boundaries. Bulk stabilizers are arranged as translational copies of a fundamental stabilizer tile on a two-dimensional plane, while boundary stabilizers are defined by restricting the support of bulk stabilizers such that all stabilizers mutually commute.

The bulk stabilizers admit a compact representation in terms of polynomials of the stabilizer supports. This polynomial formalism was originally introduced for general qudit stabilizer codes by Haah \cite{haah_commuting_2013,Haah_2017}, and was subsequently adapted to planar quantum LDPC constructions. In the present work, we specialize this formalism to two-level systems (qubits) and use it to describe the structure of bulk stabilizers in tile codes, using a similar convention to that of \cite{Liang_notation_2025}.

\begin{table*}
\centering
\caption{Planar qLDPC code families used in this paper, which are introduced in Ref.~\cite{liang_planar_2025}. The qubit numbers $n$ reported here are the numbers after grafting (see main text).}
\label{tab:tile_code_families}
\begin{tabular}{ccccccc}
\hline
Code family & $f(x,y)$ & $g(x,y)$ & $n$ & $(L_x,L_y)$ & $[[n,k,d]]$ & efficiency $\frac{k d^2}{n}$ \\
\hline
\hline

\multirow{4}{*}{$[[288,8,12]]$ family}
& \multirow{4}{*}{$1+x+x^{-1}y^{2}$}
& \multirow{4}{*}{$1+y+x^{-2}y^{-1}$}
& \multirow{4}{*}{$2L_xL_y$}
& $(6,6)$   & $[[72,8,4]]$    & $1.78$ \\
&  &  &  & $(8,8)$   & $[[128,8,6]]$  & $2.25$ \\
&  &  &  & $(10,10)$ & $[[200,8,9]]$  & $3.24$ \\
&  &  &  & $(12,12)$ & $[[288,8,12]]$ & $4.00$ \\
\hline

\multirow{6}{*}{$[[188,8,9]]$ family}
& \multirow{6}{*}{$1+x+x^{-1}y^{2}$}
& \multirow{6}{*}{$1+y+xy^{3}$}
& \multirow{6}{*}{$2L_xL_y-L_x-L_y+1$}
& $(6,10)$  & $[[105,8,6]]$  & $2.74$ \\
&  &  &  & $(7,11)$  & $[[137,8,7]]$  & $2.86$ \\
&  &  &  & $(8,12)$  & $[[173,8,8]]$  & $2.96$ \\
&  &  &  & $(8,13)$  & $[[188,8,9]]$  & $3.45$ \\
&  &  &  & $(9,14)$  & $[[230,8,10]]$ & $3.48$ \\
&  &  &  & $(10,15)$ & $[[276,8,11]]$ & $3.51$ \\
\hline

\multirow{5}{*}{$[[131,7,7]]$ family}
& \multirow{5}{*}{$1+x+x^{-1}y$}
& \multirow{5}{*}{$1+y+xy^{3}$}
& \multirow{5}{*}{$2L_xL_y-2L_x-L_y+2$}
& $(7,11)$  & $[[131,7,7]]$  & $2.62$ \\
&  &  &  & $(8,12)$  & $[[166,7,8]]$  & $2.70$ \\
&  &  &  & $(9,13)$  & $[[205,7,10]]$ & $3.41$ \\
&  &  &  & $(10,14)$ & $[[248,7,11]]$ & $3.42$ \\
&  &  &  & $(10,15)$ & $[[267,7,12]]$ & $3.78$ \\
\hline

\multirow{5}{*}{$[[88,6,6]]$ family}
& \multirow{5}{*}{$1+x+x^{-1}y^{-2}$}
& \multirow{5}{*}{$1+y+xy^{-1}$}
& \multirow{5}{*}{$2L_xL_y-L_y$}
& $(6,8)$  & $[[88,6,6]]$   & $2.45$ \\
&  &  &  & $(8,10)$ & $[[150,6,8]]$  & $2.56$ \\
&  &  &  & $(8,11)$ & $[[165,6,9]]$  & $2.95$ \\
&  &  &  & $(10,12)$& $[[228,6,11]]$ & $3.18$ \\
&  &  &  & $(10,13)$& $[[247,6,12]]$ & $3.50$ \\
\hline

\end{tabular}
\end{table*}

\textbf{Polynomial Representation.} 
We consider an open-boundary 2D lattice of size $L_x \times L_y$. 
Qubits are placed on edges, with horizontal edges defined as Left (\texttt{L}) qubits and vertical edges defined as Right (\texttt{R}) qubits. 

For algebraic convenience, we define the stabilizer structure over the Laurent polynomial ring $R = \mathbb{F}_2[x^{\pm1},y^{\pm1}]$,
where conjugation is given by 
$\bar{h}(x,y) = x^{-1}y^{-1}h(x^{-1},y^{-1})$.
Open boundary conditions are obtained by restricting to a finite region $0 \le i < L_x$, $0 \le j < L_y$. 

Each bulk stabilizer is represented as an element of the $R$-module $R^2$, where the two components correspond to the supports on L and R qubits, respectively. 
Such a stabilizer is specified by a polynomial pair $(f,g) \in R^2$:
\begin{align}
f(x,y) &= 1 + x + x^a y^b, \\
g(x,y) &= 1 + y + x^c y^d,
\end{align}
where $(a,b)$ and $(c,d)$ are integer parameters determining the code structure. 
The $X$-type stabilizers correspond to the pair $(f,g)$, whereas the $Z$-type stabilizers correspond to the conjugate pair $(\bar{g}, \bar{f})$.

For implementation, we normalize the polynomials so that all exponents are non-negative. 
Specifically, we define $f_{\text{norm}}(x,y) = x^{\alpha} y^{\beta} f(x,y)$ and 
$g_{\text{norm}}(x,y) = x^{\gamma} y^{\delta} g(x,y)$, where $\alpha = -\min\{0,a\}$, $\beta = -\min\{0,b\}$, $\gamma = -\min\{0,c\}$, and $\delta = -\min\{0,d\}$. 
Each monomial term with coefficient $1$ in $f_{\text{norm}}$ or $g_{\text{norm}}$ corresponds to a qubit in the stabilizer support.
In particular, for the $X$-type stabilizer $(f_{\text{norm}}, g_{\text{norm}})$, a term $x^i y^j$ in $f_{\text{norm}}$ specifies an \texttt{L} qubit located at position $(i,j)$ relative to the stabilizer anchor in the horizontal edge layer, while a term $x^i y^j$ in $g_{\text{norm}}$ specifies an \texttt{R} qubit at the same relative position in the vertical edge layer.

Fig.~\ref{fig:tilecode_structure} illustrates this construction for the [[288, 8, 12]] code family with parameters $(a,b,c,d) = (-1, 2, -2, -1)$ on a $L_x \times L_y = 6 \times 6$ lattice with distance $d=4$. The normalization yields $f_{\text{norm}} = x + x^2 + y^2$ and $g_{\text{norm}} = x^2y + x^2y^2 + 1$.
For each $X$-type check operator positioned at lattice site $(x_0, y_0)$, the support consists of \texttt{L} qubits at positions corresponding to the three terms in $f_{\text{norm}}$ and \texttt{R} qubits at positions corresponding to the three terms in $g_{\text{norm}}$, resulting in weight-6 bulk stabilizers. 
The highlighted bulk stabilizer demonstrates this structure: the three \texttt{L} qubits (orange) correspond to the terms $x$, $x^2$, and $y^2$ in $f_{\text{norm}}$, positioned at relative coordinates $(1,0)$, $(2,0)$, and $(0,2)$ from the check operator, while the three \texttt{R} qubits (green) correspond to the terms $x^2y$, $x^2y^2$, and $1$ in $g_{\text{norm}}$, positioned at $(2,1)$, $(2,2)$, and $(0,0)$ respectively. The constant term $1$ in $g_{\text{norm}}$ marks the origin of the support region, indicated by the X marker in the figure.
Boundary stabilizers have reduced weight due to the finite lattice size, as shown by the weight-4 boundary stabilizer, where some polynomial terms fall outside the lattice boundaries.
The $Z$-type stabilizers follow the same geometric structure but with the conjugate polynomial pair $(\bar{g}_{\text{norm}}, \bar{f}_{\text{norm}})$, ensuring the commutation relations required for a valid CSS code.

We summarize the properties of the four tile code families and their specific realization within certain lattice sizes in Table~\ref{tab:tile_code_families}.

\textbf{Code Grafting.} Conceptually, each layer forms an $L_x \times L_y$ array, so the (ungrafted) code has $2L_xL_y$ data qubits. However, as proposed in~\cite{liang_planar_2025}, one can remove unnecessary qubits: specifically, the \texttt{R} qubits on the bottom vertical-edge boundary and the \texttt{L} qubits on the rightmost horizontal-edge boundary can be removed. We remove the same number of boundary lines as proposed in~\cite{liang_planar_2025}.
The qubit numbers $n$ reported in Table~\ref{tab:tile_code_families} are the numbers after applying this grafting method.

\subsection{Noise models}

\begin{table}
\centering
\caption{Gate-dependent Pauli noise channels in the standard noise model.
Two-qubit gates (\texttt{CNOT} and \texttt{SWAP}) are each assigned the same error rate $p$.
For $\texttt{DEPOLARIZE1}(p)$, one of $\{X,Y,Z\}$ is applied with probability $p/3$.
For $\texttt{DEPOLARIZE2}(p)$, one of the $15$ non-identity two-qubit Pauli operators
is applied uniformly with probability $p/15$.}
\label{tab:noisemodel_standard}
\begin{tabular}{lll}
\hline
\textbf{Operation} & \textbf{Noise location} & \textbf{Error channel} \\
\hline
\texttt{InitX}
& after gate 
& $\texttt{Z\_ERROR}(p)$ \\

\texttt{InitZ}
& after gate 
& $\texttt{X\_ERROR}(p)$ \\

\texttt{MeasX}
& before measurement 
& $\texttt{Z\_ERROR}(p)$ \\

\texttt{MeasZ}
& before measurement 
& $\texttt{X\_ERROR}(p)$ \\

\texttt{Idle} 
& after Idle 
& $\texttt{DEPOLARIZE1}(p)$ \\

2Q Gates
& after gate 
& $\texttt{DEPOLARIZE2}(p)$ \\

\hline
\end{tabular}
\end{table}

\textbf{Standard model.}
We follow the \emph{standard circuit-based depolarizing noise model}, in the same convention as Bravyi \textit{et al.}~\cite{bravyi24high}.
Each elementary operation is assumed to undergo either a Pauli flip error or a depolarizing error with probability $p$.
Specifically, error of idle (\texttt{Idle}) operations and two-qubit gates are modeled using \texttt{DEPOLARIZE1}$(p)$ and \texttt{DEPOLARIZE2}$(p)$, respectively.
Idle noise is applied to every qubit that is idle in a given time step.
In our open-boundary circuits, this includes the ancilla qubits of reduced-weight boundary checks, which are idle during some \texttt{CNOT} ticks. The periodic-boundary circuit of Ref.~\cite{bravyi24high} has no idle ancillae, so this is where our noise placement goes beyond theirs.
For simplicity, we treat \texttt{SWAP} as a native gate with the same error rate $p$ as \texttt{CNOT}.
A detailed specification of the error model is provided in Table~\ref{tab:noisemodel_standard}.

\begin{table}[t]
\centering
\caption{Gate-dependent Pauli noise channels in the SI1000 noise model~\cite{gidney_honeycomb_2021},
with a native \texttt{SWAP} gate following Ref.~\cite{zhou_louvre_2025}.
The base error probability $p$ is the two-qubit \texttt{CNOT} gate error rate.
Gate idle ($p/10$) acts on all inactive qubits every tick; resonator idle ($2p$)
acts additionally on non-M/R qubits during ticks that contain a measurement or
reset~\cite{gidney_honeycomb_2021}. Both idle channels stack on M/R ticks.
Abbreviations: \texttt{DEP1} = \texttt{DEPOLARIZE1},
\texttt{DEP2} = \texttt{DEPOLARIZE2}.}
\label{tab:noisemodel_si1000}
\begin{tabular}{lll}
\hline
\textbf{Operation} & \textbf{Noise location} & \textbf{Channel} \\
\hline
\texttt{Init}       & after reset              & \texttt{X\_ERROR}$(2p)$ \\
1Q gates            & after gate               & \texttt{DEP1}$(p/10)$ \\
\texttt{CNOT}   & after gate               & \texttt{DEP2}$(p)$ \\
\texttt{SWAP}       & after gate               & \texttt{DEP2}$(1.5p)$ \\
\texttt{Meas}       & before meas.             & \texttt{X\_ERROR}$(5p)$ \\
Idle (gate)         & every tick, inactive     & \texttt{DEP1}$(p/10)$ \\
Idle (resonator)    & M/R ticks, non-M/R       & \texttt{DEP1}$(2p)$ \\
\hline
\end{tabular}
\end{table}

\textbf{SI1000 model.}
In addition to the standard model, we also simulate a \textit{modified} version of the \textit{SI1000 noise model}~\cite{gidney_honeycomb_2021}, which is designed to match the noise profile of superconducting transmon processors.
The model uses a single parameter $p$ set by the two-qubit gate error rate, and all other error rates are fixed multiples of $p$. A full specification is given in Table~\ref{tab:noisemodel_si1000}. Unlike the original SI1000 definition, which uses \texttt{CZ} as the native two-qubit gate, we treat \texttt{CNOT} as the native two-qubit gate, since our syndrome extraction circuit is \texttt{CNOT}-based.

The SI1000 model has two separate idle error channels that are both applied independently~\cite{gidney_honeycomb_2021}. The first is a gate idle of $\texttt{DEPOLARIZE1}(p/10)$, applied to qubits not doing anything in each cycle. The second is a resonator idle of $\texttt{DEPOLARIZE1}(2p)$, applied in cycles that contain a measurement or reset, to qubits not being measured or reset. Both channels apply at the same time when a qubit is idle during a measurement or reset cycle.

For the routing simulations that we describe below, we assign the \texttt{SWAP} gate an error rate of $1.5p$, following Ref.~\cite{zhou_louvre_2025}. On tunable-coupler superconducting hardware, a \texttt{SWAP} pulse is about $1.5\times$ noisier than a \texttt{CZ} or \texttt{iSWAP} gate when decoherence is the main source of error~\cite{zhou_louvre_2025}. This lets us treat \texttt{SWAP} as a single native gate rather than decomposing it into three \texttt{CZ} gates.

\subsection{Logical error rates in circuit-level noise model}\label{sec:pl_def}
In order to capture the effect of circuit-level noise over time, the syndrome measurement cycle is repeated for $N$ rounds. Following the standard simulation protocol~\cite{tomita2014low,bravyi24high}, we choose $N=d$, where $d$ denotes the code distance. This choice provides a natural benchmark for evaluating logical error rates in a fault-tolerant memory setting.

The accumulated logical error probability after $N$ syndrome measurement cycles can be converted into an effective logical error rate per cycle. Denoting by $p_{L,N}$ the probability of at least one logical failure over $N$ cycles, we define the per-cycle logical error rate as
\[
p_L = 1 - (1 - p_{L,N})^{1/N}.
\]

\section{Results}
\subsection{Scheduling of syndrome measurement circuit} \label{section:scheduling_swap}
\textbf{Without locality constraints.}
When two-qubit gates can be applied between an ancilla and the corresponding data qubit without locality constraints, the scheduling strategy of the BB code~\cite{bravyi24high} that performs syndrome measurements of \texttt{L} and \texttt{R} qubits in parallel to minimize circuit depth also works for tile codes.
In this case, the SI1000 model uses the same syndrome measurement cycle as the standard model; the two models differ only in their error rates. Our eight-round schedule achieves the minimum depth consistent with weight-6 stabilizers: each check operator requires at least one initialization, six \texttt{CNOT} operations (one per support qubit), and one measurement round. Open-boundary stabilizers have reduced weight and are accommodated within the same schedule without additional overhead~\cite{zhang_asc_2026}.

\begin{table}[t]
\centering
\caption{Optimal routing schedules for each tile-code family. $L$: number of \texttt{SWAP} steps per round. CX ticks: parallel \texttt{CNOT} ticks after packing. Total = $L$ + CX ticks. $\delta_Z$, $\delta_X$: starting offsets of $Z$- and $X$-ancillae within the stabilizer tile.}
\label{tab:routing_schedules}
\begin{tabular}{lcccccc}
\hline
Family & Word $\mathcal{W}$ & $L$ & CX ticks & Total & $\delta_Z$ & $\delta_X$ \\
\hline
$[[288,8,12]]$ & \texttt{RRDLDRR}  & 7 & 8  & 15 & $(-1,0)$ & $(0,0)$ \\
$[[88,6,6]]$   & \texttt{LLDRDLL}  & 7 & 8  & 15 & $(2,0)$  & $(3,0)$ \\
$[[131,7,7]]$  & \texttt{RRDDLDRR} & 8 & 10 & 18 & $(-1,0)$ & $(0,0)$ \\
$[[188,8,9]]$  & \texttt{RRDDLDRR} & 8 & 10 & 18 & $(-1,0)$ & $(0,0)$ \\
\hline
\end{tabular}
\end{table}

\textbf{Routing schemes.}
We next consider the case where two-qubit gates are only allowed between nearest-neighbor qubits on a two-dimensional planar lattice. 
In this case, measuring nonlocal stabilizers requires qubit routing, i.e., an ancilla qubit should move around the lattice to locate the corresponding data qubit to be its nearest neighbor.
This motivates a general \texttt{SWAP}-based routing scheme applicable to any planar tile-code family.

We represent a routing strategy as a \emph{routing word} $\mathcal{W} = w_1 w_2 \cdots w_L$ with $w_i \in \{\mathrm{L}, \mathrm{R}, \mathrm{U}, \mathrm{D}\}$.
Each letter specifies the direction of one \texttt{SWAP} step; all ancillae follow the same word, but $Z$- and $X$-ancillae may start from different lattice offsets ($\delta_Z$ and $\delta_X$, respectively) relative to the data qubits of the stabilizer they measure. For each swap, data qubits corresponding to each swapped ancillae move to the opposite direction.
Between consecutive \texttt{SWAP} steps, each ancilla applies \texttt{CNOT} gates to adjacent support qubits. 
For the even-numbered rounds, we execute the time-reverse of the forward schedule and ensure the ancilla returns to its initial position.
The total circuit depth per round is $L$ \texttt{SWAP} steps plus the number of \texttt{CNOT} sub-ticks.

We find valid routing words through a three-phase exhaustive search.

\textit{Phase~1 (Coverage).}
For each candidate word $\mathcal{W}$ and ancilla starting offset, we check that the ancilla visits all support qubits of every stabilizer during the sweep. Boundary stabilizers with fewer than six support qubits are handled automatically.

\textit{Phase~2 (CNOT ordering).}
For each word that passes Phase~1, we enumerate all valid assignments of support qubits to \texttt{SWAP} slots and \texttt{CNOT} orderings within each slot. A schedule is accepted if, for every pair of overlapping Z and X stabilizers, the number of shared qubits on which the X \texttt{CNOT} fires before the Z \texttt{CNOT} is even. This is condition~(b') of~\cite{geher2024tangling}, which guarantees that both stabilizer types are measured simultaneously and independently.

\begin{remark}[CSS specialization of condition (b')]
For CSS codes, anti-commuting Pauli pairs in the sense of~\cite{geher2024tangling} can only arise between X-type and Z-type check operators. Condition~(b') therefore reduces to: for every pair $(S_X, S_Z)$ of stabilizers with overlapping support, the number of shared data qubits on which the X-\texttt{CNOT} fires before the Z-\texttt{CNOT} must be even. This is precisely the criterion enforced in Phase~2 of our exhaustive search.
\end{remark}

\textit{Phase~3 (Sub-tick packing).}
\texttt{CNOT} tokens within a slot are merged into parallel ticks whenever they act on disjoint qubits, minimizing the total number of \texttt{CNOT} sub-ticks per round.

Finally, we allow the $Z$- and $X$-ancillae to use independently chosen starting offsets (dual-offset search), which yields shorter valid words for three of the four families.
The resulting optimal schedules are listed in Table~\ref{tab:routing_schedules} and illustrated in Appendix~\ref{sec:appendix_routing}. 

We verify that each ancilla interacts with every qubit in its support exactly once per round, and that the syndrome-extraction circuit produces a deterministic detector error model.
Also, since each odd-numbered round is followed by its time reverse, all SWAP operations cancel out and every data qubit returns to its original position after two rounds, so no frame tracking is needed.

While our exhaustive search algorithm here is developed around the tile code families, it can be applied to other transversal-invariant codes.

\subsection{Circuit-level thresholds}
\textbf{Estimation method.}
We simulated our syndrome-measurement circuits using the Stim circuit simulator~\cite{stim} with the BP+OSD decoder. 
For each physical error rate, we ran the circuits $5,000-70,000$ times to estimate the logical error rate $p_L$ and estimated the threshold $p_{\rm th}$ using the finite-size scaling method.
To reduce statistical error, we collected at least 50 logical errors for each $(p, d)$ point included in the finite-size scaling fit.
Precisely, we fit the logical error rate to the quadratic ansatz $p_L = A + B x + C x^2$ with $x = (p - p_{\rm th})\, d^{1/\nu}$, where $p_{\rm th}$ is the threshold, $\nu$ is the correlation-length exponent,
and $A$, $B$, $C$ are fitting coefficients.
We treat $\nu$ as an effective fitting parameter rather than a universal critical exponent,
since the code distance does not scale linearly with the lattice dimensions. Since $\nu$ is expected to differ across families and code distances in this finite-size regime, we treat it as a family- and model-specific fit parameter.

The threshold is determined self-consistently by iterating the fit window until $p_{\rm th}$ converges; the window is then held fixed across all bootstrap resamples.
The fit uses data points within the $p$ window $[0.5\,p_{\rm th},\, 2.0\,p_{\rm th}]$, and we further restrict to $p_L \in [0.03, 0.97]$ (no-routing) or $p_L \in [0.005, 0.60]$ (routing), to avoid saturated or near-zero logical error rates where the ansatz breaks down.

\begin{table}
\centering
\caption{Circuit-level error thresholds for each tile-code family, estimated from a quadratic
finite-size scaling fit $p_L = A + Bx + Cx^2$, $x=(p-p_{\rm th})\,d^{1/\nu}$.
The thresholds $p_{\rm th}$ and $p_{\rm th,SI}$ are obtained under the standard and SI1000 noise models without locality constraints, whereas $p_{{\rm th},r}$ and $p_{{\rm th,SI},r}$ are the corresponding values with \texttt{SWAP} routing.
Parenthetical uncertainties are bootstrap standard deviations ($N=300$ resamples).
}
\begin{tabular*}{\columnwidth}{@{\extracolsep{\fill}}lcccc}
\hline
Family & $p_{\rm th}$ & $p_{\rm th,SI}$ & $p_{{\rm th},r}$ & $p_{{\rm th,SI},r}$ \\
\hline
$[[288,8,12]]$ & 0.527(35)\% & {0.306(11)}\% & {0.168(7)}\% & {0.134(3)}\% \\
$[[188,8,9]]$  & {0.503(39)}\% & {0.295(25)}\% & {0.147(3)}\% & {0.119(3)}\% \\
$[[131,7,7]]$  & {0.437(31)}\% & {0.234(8)}\%  & {0.136(6)}\% & {0.110(4)}\% \\
$[[88,6,6]]$   & {0.476(16)}\% & {0.259(24)}\% & {0.155(8)}\% & {0.117(4)}\% \\
\hline
\end{tabular*}
\label{tab:fg_families}
\end{table}

\begin{figure*}[t]\centering
\includegraphics[width=\linewidth]{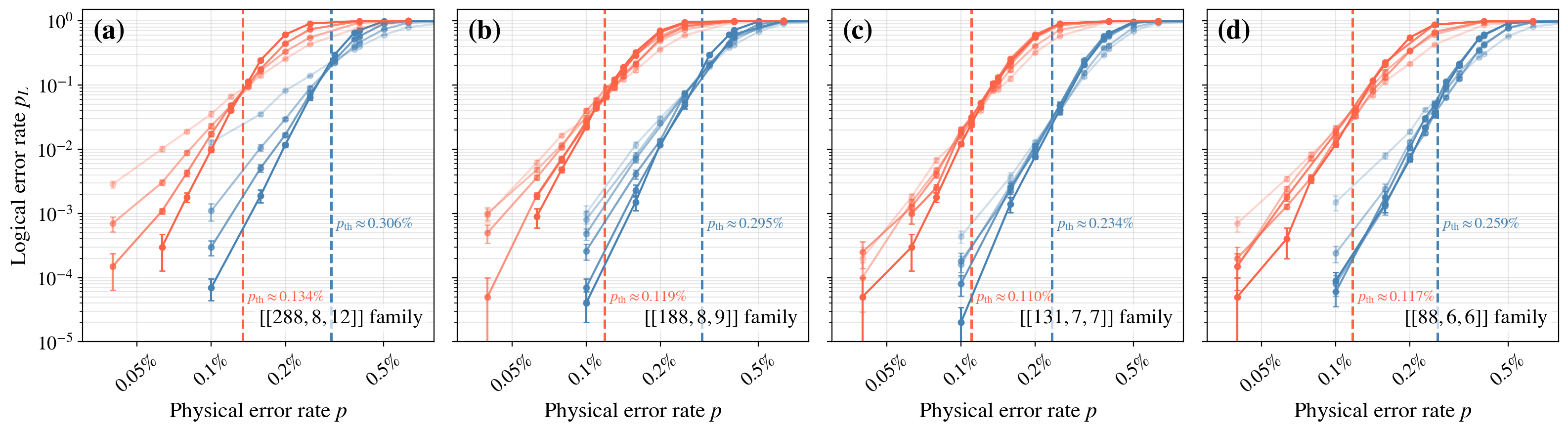}
\caption{
Logical error rate $p_L$ versus physical error rate $p$ for the (a) $[[288,8,12]]$, (b) $[[188,8,9]]$, (c) $[[131,7,7]]$, and (d) $[[88,6,6]]$ code families under the SI1000 noise model.
Blue curves correspond to results without locality constraints, and red curves to results with \texttt{SWAP} routing, where two-qubit gates are only allowed between nearest-neighbor qubits.
Within each panel, lighter shades indicate smaller code distances (see Table~\ref{tab:tile_code_families}). Dashed vertical lines indicate the threshold for each family obtained from the finite-size scaling.
}
\label{fig:pl_si1000}
\end{figure*}

\textbf{Threshold under the circuit-level noise models.}
We plot the results for the SI1000 noise model with and without locality constraints in Fig.~\ref{fig:pl_si1000}. 
In our simulation, we decoded only the syndromes of the $Z$-stabilizer, i.e., we corrected $X$-errors, and all reported thresholds refer to this decoding problem.
This is the standard convention for CSS codes under circuit-level noise~\cite{hillmann_lsd_2024, gong_sliding_2024, li_dimension_jump_2025}; for BP-based decoders, restricting to a single detector basis has moreover been observed to \textit{improve} decoding accuracy by avoiding trapping sets induced by $Y$-type errors~\cite{beni_tesseract_2025}.
We expect the $X$-error threshold to coincide with the reported values: every noise channel in both models is invariant under a transversal-Hadamard change of frame, under which our \texttt{CNOT}-based syndrome-extraction circuit maps exactly onto an $X$-basis memory circuit of the boundary-dual code, and the bulk stabilizer structure of tile codes is symmetric under exchanging $X$- and $Z$-type checks (we verified numerically that the periodic-boundary code defined by the same polynomials $f$ and $g$ has isomorphic $X$ and $Z$ Tanner graphs for all four families).
The two decoding problems therefore differ only near the open boundaries, which affect sub-threshold logical error rates but not the threshold itself.
We provide chosen parameters for BP+OSD and a comprehensive performance comparison with other decoders in Appendix~\ref{section:decoder}.

For the [[288,8,12]] code family, the threshold under the SI1000 noise model without locality constraints is given as $p_{\rm th,SI}\approx 0.306\%$ whereas it reduces to $p_{\rm th,SI,r}\approx 0.134\%$ with a routing scheme, where two-qubit gates are only allowed between nearest-neighbor qubits.
A similar behavior is observed in other code families, where the threshold is reduced by a factor of $2.1$ to $2.5$, indicating the cost of additional \texttt{SWAP} gates.

In Appendix~\ref{app:threshold_standard}, we also run the same calculation for the standard noise model and summarize the obtained thresholds in each setting in Table~\ref{tab:fg_families}.
We observe that the thresholds with locality constraints are reduced by a factor of $\approx 3$ for the standard noise model, which is larger than in the SI1000 case.
We believe this is because the idle error is much larger in the standard noise model.

\subsection{Resource efficiency compared to the surface code}\label{sec:footprint}

\begin{figure}[t]\centering
\includegraphics[width=\linewidth]{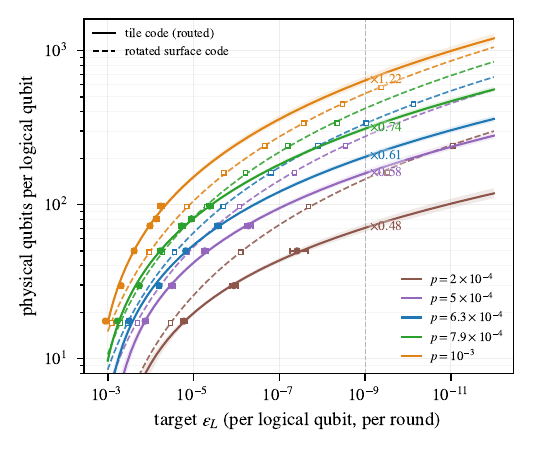}
\caption{
The number of physical qubits per logical qubit needed to reach a target error rate $\epsilon_L$ under the SI1000 model, for the routed $[[288,8,12]]$ family (solid) and the rotated surface code (dashed). Colors denote the physical error rate $p$, and bands are bootstrap $1\sigma$ ranges. Circles and squares are directly simulated instances. Labels give the tile-to-surface ratio at $\epsilon_L = 10^{-9}$ (dashed vertical line). 
}
\label{fig:footprint}
\end{figure}

The reduced thresholds above do not directly tell us the hardware cost.
A more practical measure is the number of physical qubits needed to reach a target logical error rate at a given physical error rate.
We answer it for the $[[288,8,12]]$ family with routing, our best-performing family.
We convert the measured block logical error probability $P$ into a per-logical-qubit, per-round error rate $\epsilon_L$ through $P = 1-(1-\epsilon_L)^{k r}$, with $k=8$ logical qubits and $r=d$ rounds.
Unlike the per-cycle rate $p_L$ of Sec.~\ref{sec:pl_def}, $\epsilon_L$ is also normalized per logical qubit, so codes with different $k$ can be compared fairly.

Following Ref.~\cite{gidney_planar_honeycomb_2022}, we fit $\ln \epsilon_L$ against $d$ at each fixed $p$.
The slope gives the suppression factor $\Lambda$, the factor by which $\epsilon_L$ drops when $d$ increases by two.
From the fit we obtain the distance needed to reach any target $\epsilon_L$, and from that the number of physical qubits per logical qubit, $n_{\rm total}(d)/k$. 
The baseline is the rotated surface code decoded with minimum-weight perfect matching, treated in the same way: it is simulated under the identical circuit-level noise model, including idle depolarization applied to every qubit in every gate layer, and its qubit count is taken with the same data-plus-ancilla convention as the tile codes (one ancilla per independent stabilizer check), giving $2d^2-1$ physical qubits per logical qubit.

In addition to the instances used in the threshold study, we simulated two larger family members, $[[338,8,13]]$ and $[[392,8,15]]$.
For data points at higher error rates, we collected $50$ logical errors per point besides of the largest intsance where we only collected $30$ errors.
For points deep below threshold, where logical errors become rare, we instead fixed a shot budget of up to $1.8\times10^6$ shots per point, set by the decoding cost.
On the surface-code side, we simulated up to $d=15$ and $17$ directly, so the surface baseline near $\epsilon_L = 10^{-9}$ is measured rather than extrapolated. One caveat is that tile codes are decoded with BP+OSD while the surface code uses MWPM, the standard choice for each code.

Figure~\ref{fig:footprint} shows the resulting footprints under the SI1000 model. At $p = 10^{-3}$, which is about $75\%$ of the routed threshold, the routed tile code is \textit{more} expensive than the surface code, by a factor of $1.21$ at $\epsilon_L = 10^{-9}$. As $p$ decreases, the footprint drops quickly.
The tile-to-surface physical qubit ratio at $\epsilon_L = 10^{-9}$ falls monotonically from $1.21$ at $p = 10^{-3}$ to $0.48$ at $p = 2\times10^{-4}$, crossing unity near $p^* \approx 8\times10^{-4}$, i.e., the routed tile codes become more resource-efficient than the surface code when $p<p^*$.

We also conduct the same analysis for the standard depolarizing model in Appendix~\ref{app:footprint_standard}, which shows the same trend, but the routed tile code is already ahead at $p = 10^{-3}$ for every target we consider. It needs $30\%$ fewer qubits than the surface code at $\epsilon_L = 10^{-6}$, and $41\%$ ($47\%$) fewer at $10^{-9}$ ($10^{-12}$). 
In short, once the physical error rate is well below threshold, tile codes become a more qubit-efficient planar memory even after paying the full routing cost.

\section{Discussion}
In this work, we developed a routing scheme that implement syndrome extraction for four tile-code families using only nearest-neighbor gates on a two-dimensional square lattice
and performed circuit-level simulations of two representative gate-based noise models.
Across all four families in Table~\ref{tab:tile_code_families}, the estimated thresholds under the SI1000 model range from $p_{\rm th,SI} = 0.23\%$ to $0.31\%$, indicating that tile codes can maintain a competitive circuit-level threshold while using weight-6 local stabilizers and open boundaries.
This threshold extraction is well-defined because the logical dimension $k$ is determined entirely by the bulk polynomials $f$ and $g$ and does not depend on the lattice size (in contrast to BB codes, where $k$ varies with the code dimensions~\cite{bravyi24high}).
When \texttt{SWAP} routing is included, thresholds are reduced to $p_{{\rm th,SI},r} = 0.11\%$--$0.13\%$ across all four families, demonstrating that connectivity constraints and additional two-qubit operations can substantially reshape the effective operating regime on planar hardware.

The routing finding algorithm can be a standalone tool.
Our three-phase search schedules syndrome-measurement circuits on a strictly nearest-neighbor square grid with every detector deterministic.
Our algorithm only relies on the translation invariance of the stabilizers, so the same procedure carries over to other translation-invariant families, to other lattice geometries, and even to codes defined in three dimensions in principle.
Because our search is exhaustive over the routing-word class, the reported tick counts (15–18 per round) are optimal within that class.

With explicit schedules in hand, the price of locality can be measured rather than assumed. 
Enforcing nearest-neighbor connectivity lowers the threshold by a factor of two to three, consistently across all four families and both noise models.
A lower threshold, however, is not the final verdict, because tile codes buy back qubits through their encoding rate. The footprint metric of Sec.~\ref{sec:footprint} weighs both effects in a single number. Under the standard model the routed tile code is already ahead at $p = 10^{-3}$ for every target we consider, and under SI1000 the advantage sets in below $p^* \approx 8\times10^{-4}$ and grows as the physical error rate decreases.

We note a few limitations of the present study.
First, the routing schedules reported here are optimal within the class of routing words considered (uniform \texttt{SWAP} direction per step, translation-invariant assignment). Further improvement may be possible through non-uniform protocols or co-optimized qubit placement.
Second, our decoding results rely on BP+OSD with fixed hyperparameters; different decoders or additional decoder tuning may change quantitative threshold estimates, particularly in the presence of biased or operation-dependent noise.

\textit{Note added.} While finalizing this manuscript, three closely related works~\cite{gu_nn_2026,nixon2026vine,choe_barbell_2026} are uploaded to arXiv.
\citet{gu_nn_2026} studied \textit{directional tile codes} whose stabilizer supports form a single connected path on the lattice and demonstrated constant-depth \texttt{CXSWAP}-walk measurement on a standard planar grid. 
\citet{nixon2026vine} provided a similar construction, Vine codes, using \texttt{iSWAP} and \texttt{CZ} as native gates. 
As discussed in Sec.~\ref{section:scheduling_swap}, the four families considered in this work are non-directional and are not amenable to such walks; the codes studied here and Ref.~\cite{gu_nn_2026,nixon2026vine} do not overlap. Moreover, Ref.~\cite{gu_nn_2026} reports direct simulations at a single physical error rate, whereas we additionally provide teraquop-style resource extrapolations (Sec.~\ref{sec:footprint}).
Also, we provide threshold estimation for code families, which is not available in Refs.~\cite{gu_nn_2026,nixon2026vine}.
Separately, \citet{choe_barbell_2026} introduced Barbell codes, designed for dedicated six-qubit star lattice (6QSL+NLC) chips with twelve-neighbor fixed connectivity, a hardware setting distinct from the standard planar nearest-neighbor grid targeted here.

\section*{Acknowledgments}
Authors thank Seok-Hyung Lee, Vincent Steffan, Shin Ho Choe for helpful discussion.
This research was supported by the Yonsei University Research Fund of 2024-22-0501; the BK21FOUR (Fostering Outstanding Universities for Research), funded by the Ministry of Education (MOE) of Korea and the National Research Foundation (NRF) of Korea; and the NRF grants funded by the Korean government (MSIT) (No. RS-2025-16066935, RS-2025-02316431, and  RS-2025-18362970).
This research used resources from the National Energy Research Scientific Computing Center, which is supported by the Office of Science of the U.S. Department of Energy under Contract No. DE-AC02-05CH11231 using NERSC award NERSC DDR-ERCAP0029552.

\bibliography{references}

\clearpage
\newpage

\appendix
\onecolumngrid

\section{Comparison between various decoders}\label{section:decoder}

We compare four decoding algorithms for the $[[288,8,12]]$ tile code under the standard noise model at distances $d=4$ and $d=9$: BP+OSD~\cite{roffe_decoding_2020}, BP+LSD~\cite{hillmann_lsd_2024}, BP+UFD (BeliefFind)~\cite{BeliefFind2021}, and MWPF~\cite{wu2025-uf}. All BP-based decoders share the same front-end: min-sum belief propagation with scaling factor 0.5 and a maximum of 2000 iterations (see Appendix~\ref{sec:appendix_bposd_params}).

BP+OSD achieves the lowest logical error rate across all tested distances and error rates. The advantage grows with higher code distance and lower $p$: at $d=9$ and $p = 1.585 \times 10^{-3}$, BP+OSD is $19\times$ lower than BP+LSD and $197\times$ lower than BP+UFD. BP+UFD shows the weakest sub-threshold scaling among the four decoders.

We evaluate MWPF with $\text{cluster\_node\_limit}=500$ for $d=4$ and $\text{limit}=200$ for $d=9$. At $d=4$, MWPF achieves similar $p_L$ to BP+OSD. At $d=9$, MWPF is $10$--$32\times$ worse in $p_L$ than BP+OSD across sub-threshold error rates, and near $p = 4 \times 10^{-3}$ collapses to $p_L \approx 0.47$, effectively above threshold. While MWPF's accuracy can improve with a larger cluster limit~\cite{wu2025-uf}, this trades off against substantially increased per-shot runtime, and at the tested settings BP+OSD already requires $4$--$12\times$ less runtime while achieving markedly lower $p_L$. We therefore adopt BP+OSD as our primary decoder.

\begin{figure}[H]
    \centering
    \includegraphics[width=0.6\linewidth]{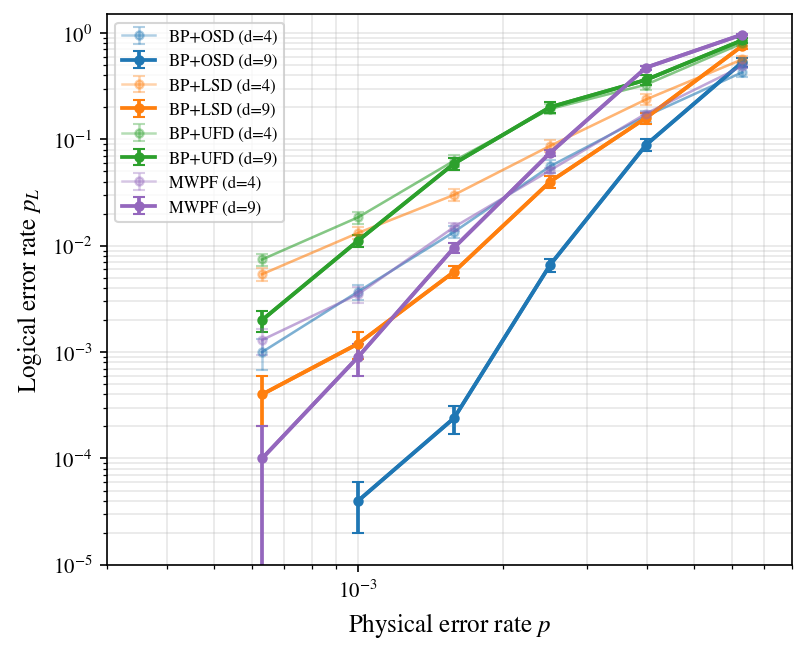}
    \caption{Logical error rate $p_L$ as a function of the physical error rate $p$ for the $[[288,8,12]]$ tile code under the standard depolarizing noise model, for code distances $d=4$ (lighter) and $d=9$ (darker). Four decoding algorithms are compared: BP+OSD, BP+LSD, BP+UFD (BeliefFind), and MWPF. All BP-based decoders use the same front-end (min-sum, scaling factor 0.5, max 2000 iterations). MWPF uses $\text{cluster\_node\_limit}=500$ for $d=4$ and $\text{limit}=200$ for $d=9$.}
    \label{fig:decoders}
\end{figure}

\clearpage
\section{BP+OSD Hyperparameter Selection}\label{sec:appendix_bposd_params}

We select the BP+OSD hyperparameters using a two-stage sweep on the no-routing $[[288,8,12]]$ circuit at $d=4$, evaluating both the standard and SI1000 noise models.

\textbf{OSD order.} We use OSD-CS with order 7. Differences between order 7 and higher orders (9, 11) are within 6\% in logical error rate, while runtime increases rapidly with order. We therefore fix order 7 throughout.

\begin{figure}[H]\centering
\includegraphics[width=\linewidth]{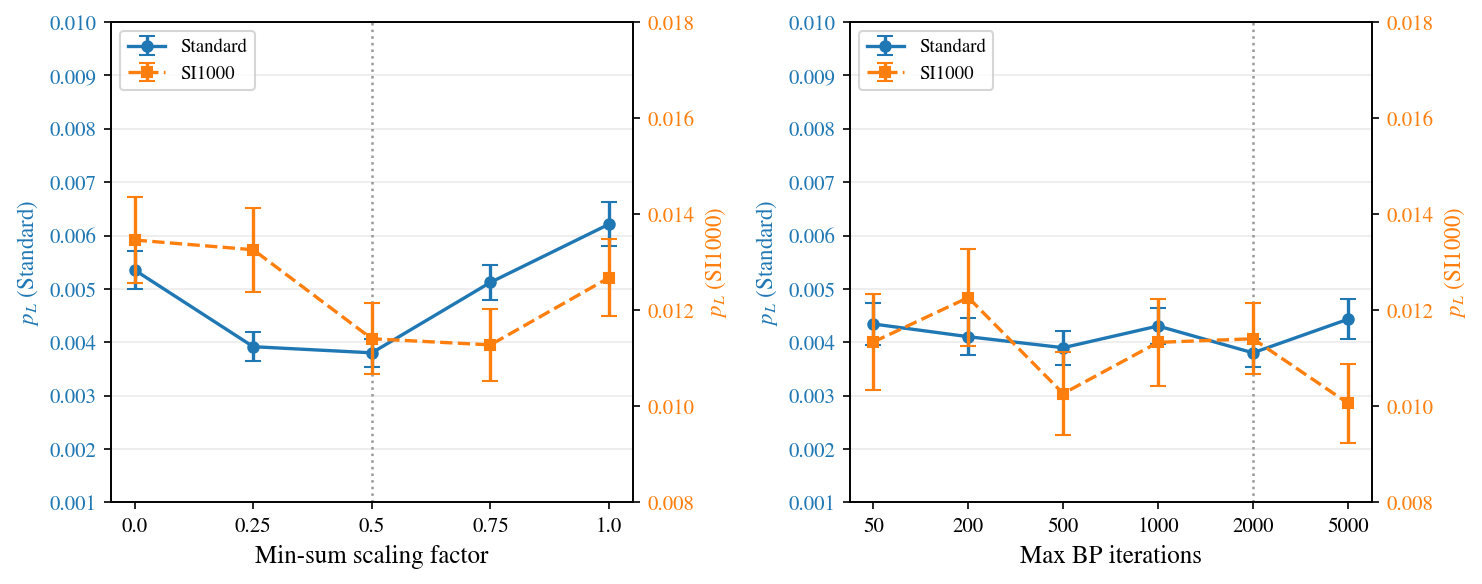}
\caption{Hyperparameter sweep for the BP+OSD decoder applied to the no-routing $[[288,8,12]]$ circuit ($d=4$) at $p=10^{-3}$ under the standard (blue, left axis) and SI1000 (orange, right axis) noise models. \textbf{Left:} logical error rate $p_L$ as a function of the min-sum scaling factor, with max iterations fixed at 2000. \textbf{Right:} $p_L$ as a function of the maximum BP iteration count, with the scaling factor fixed at 0.5. Dotted vertical lines mark the adopted values ($\text{ms}=0.5$, $\text{max\_iter}=2000$). Error bars represent Poisson statistical uncertainty.}
\label{fig:param_sweeping}
\end{figure}

\textbf{Min-sum scaling factor.} We sweep the scaling factor in $\{0, 0.25, 0.5, 0.75, 1.0\}$ with max iterations fixed at 2000 and $p = 10^{-3}$. As shown in Fig.~\ref{fig:param_sweeping} (left), $\text{ms}=0.5$ achieves the lowest or tied-lowest $p_L$ under both noise models ($+0$--$0.1\sigma$ from best). Values of 0.75 and 1.0 are significantly worse ($+3$--$5\sigma$). We fix $\text{ms}=0.5$.

\textbf{Maximum BP iterations.} We sweep max\_iter in $\{50, 200, 500, 1000, 2000, 5000\}$ with $\text{ms}=0.5$ fixed. As shown in Fig.~\ref{fig:param_sweeping} (right), $p_L$ is flat across all values for both noise models, with convergence already achieved at \texttt{max\_iter}$=50$: the no-routing circuit is shallow (TICK depth 33), so the syndrome is sparse at $p=10^{-3}$ and BP converges rapidly. OSD post-processing corrects any residual errors, making $p_L$ insensitive to the iteration budget. We conservatively adopt max\_iter$=2000$, which is also consistent with convergence verified near the threshold on the routing circuit (see the main text).

\clearpage
\section{Illustration of SWAP Routing Schedules}\label{sec:appendix_routing}
\begin{figure}[h]
    \centering
    \includegraphics[width=0.8\linewidth]{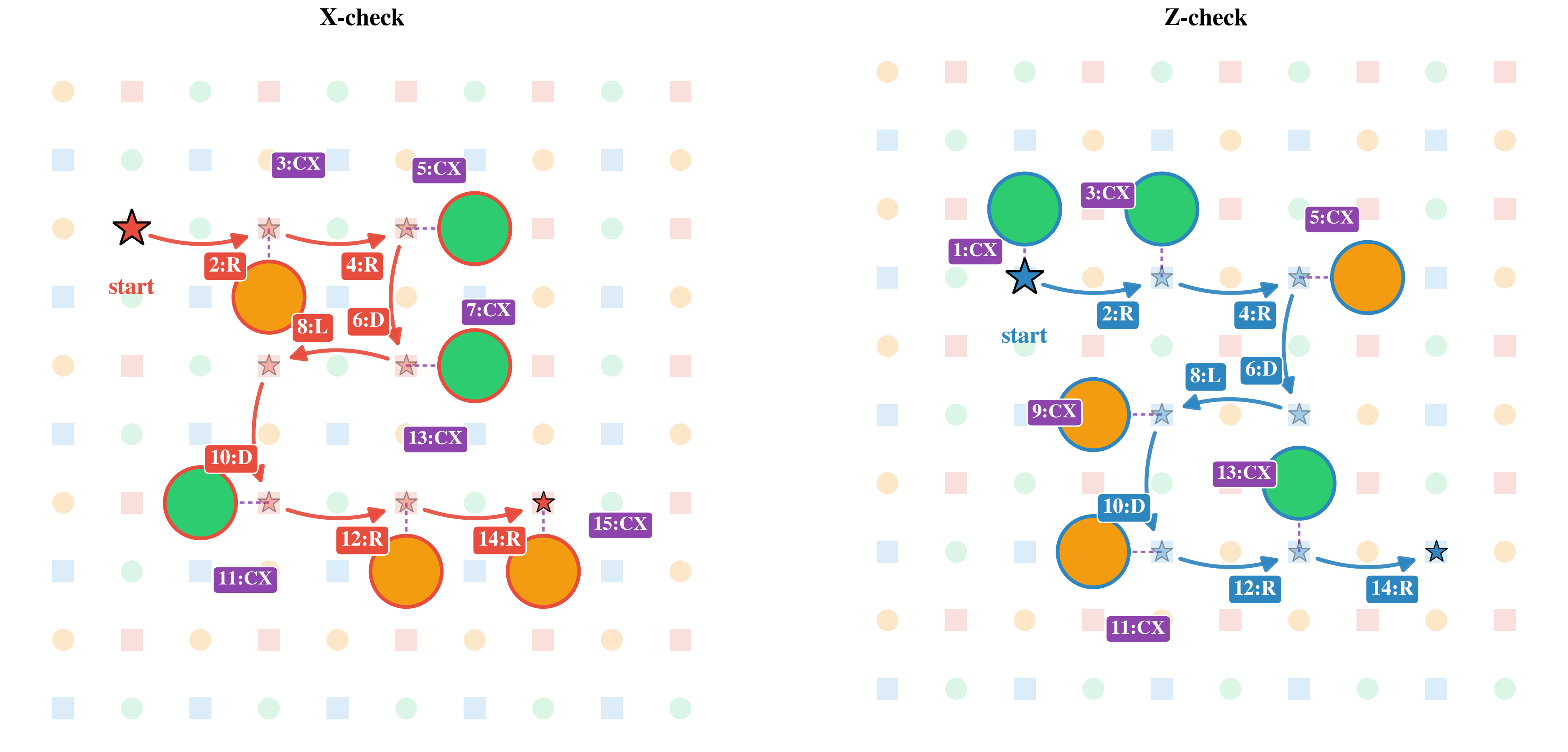}
    \caption{Routing schedule for the $[[288,8,12]]$ family. Word: \texttt{RRDLDRR}, 15 ticks/round. Green and yellow circles indicate DataR and DataL support qubits, respectively; the ancilla is shown with a distinctive marker symbol. Each tick is represented as a numbered box containing either a directional arrow (\texttt{SWAP} step) or a \texttt{CNOT} operation label.}
    \label{fig:routing_288}
\end{figure}

\begin{figure}[h]
    \centering
    \includegraphics[width=0.8\linewidth]{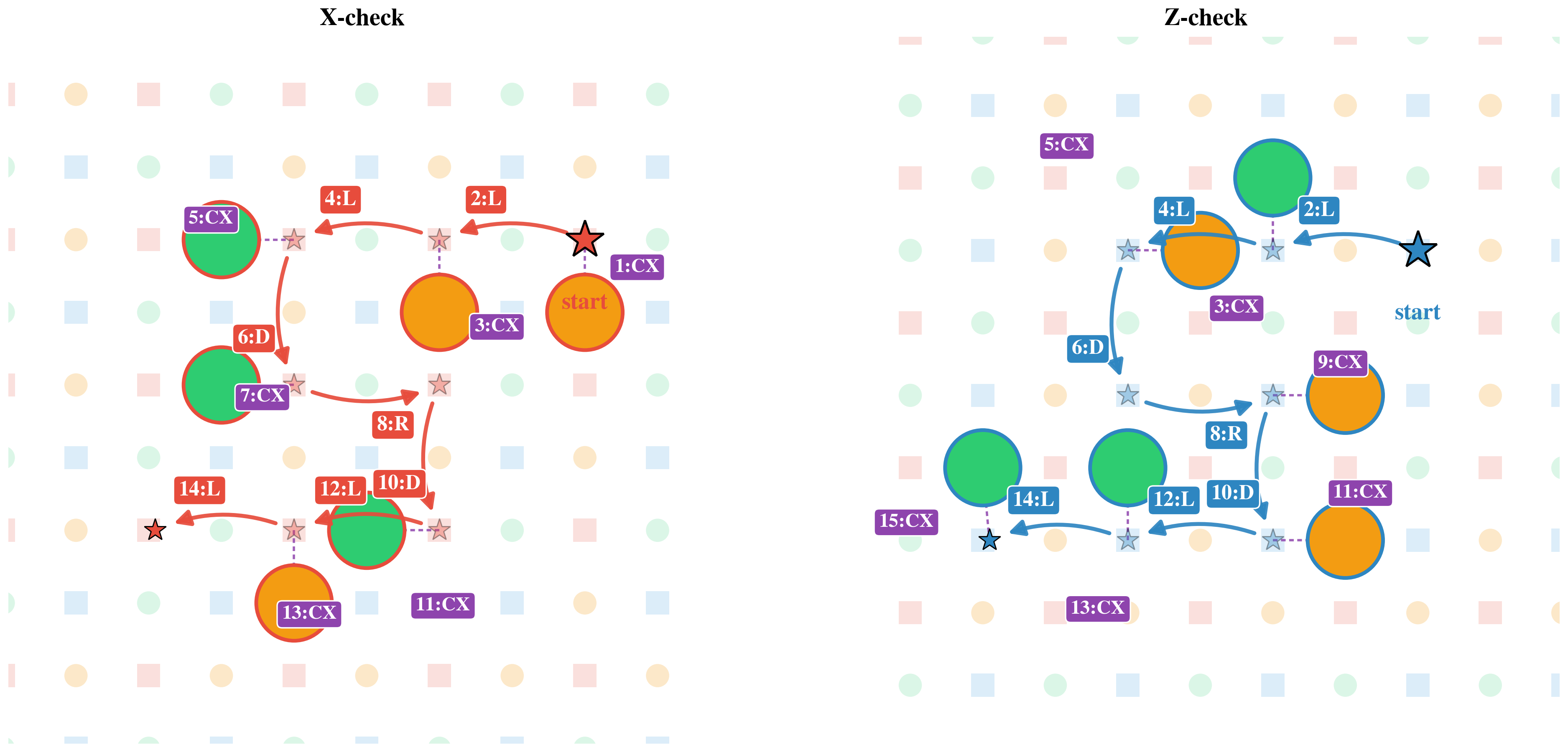}
    \caption{Routing schedule for the $[[88,6,6]]$ family. Word: \texttt{LLDRDLL}, 15 ticks/round. Green and yellow circles indicate DataR and DataL support qubits, respectively; the ancilla is shown with a distinctive marker symbol. Each tick is represented as a numbered box containing either a directional arrow (\texttt{SWAP} step) or a \texttt{CNOT} operation label.}
    \label{fig:routing_88}
\end{figure}

\begin{figure}[h]
    \centering
    \includegraphics[width=0.8\linewidth]{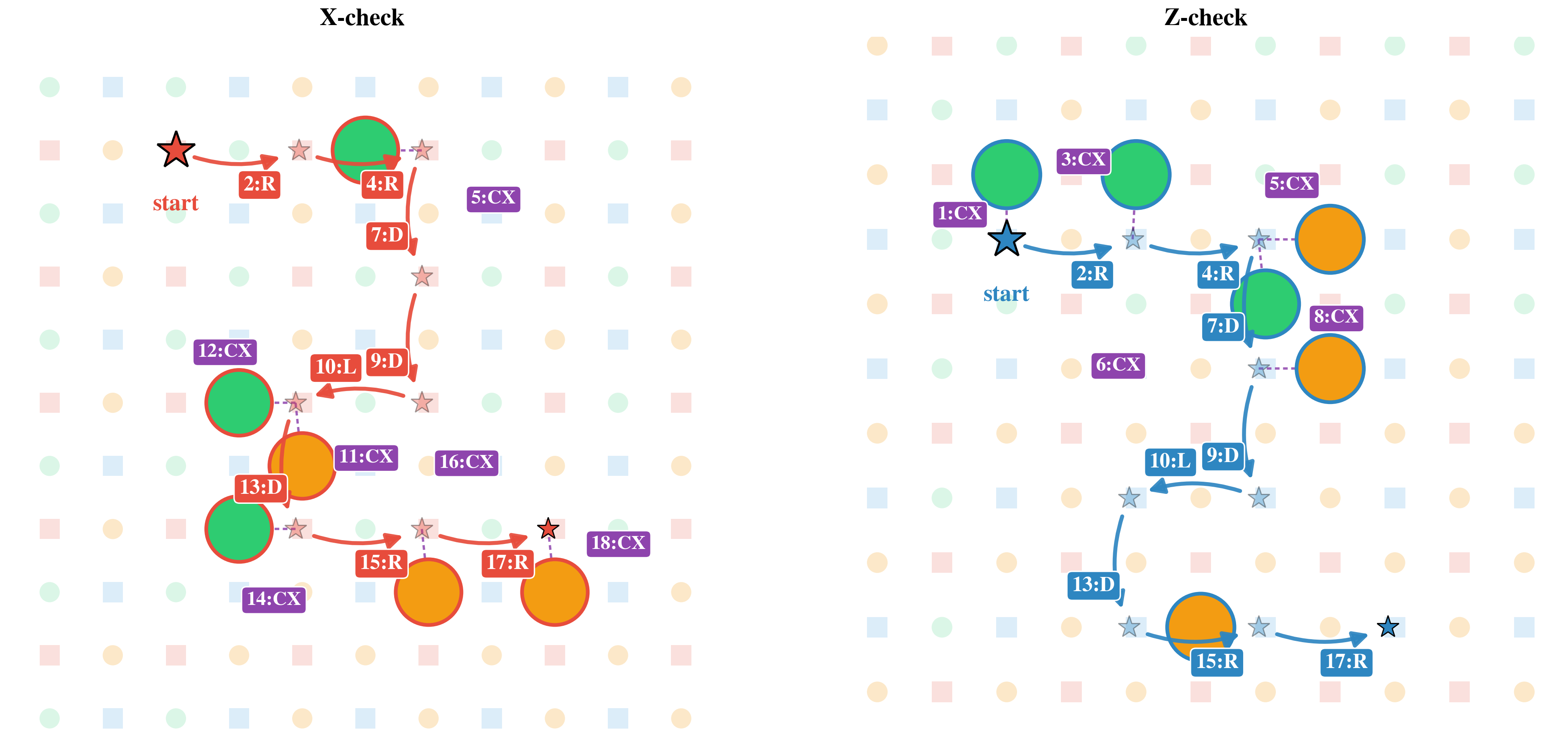}
    \caption{Routing schedule for the $[[131,7,7]]$ family. Word: \texttt{RRDDLDRR}, 18 ticks/round. Green and yellow circles indicate DataR and DataL support qubits, respectively; the ancilla is shown with a distinctive marker symbol. Each tick is represented as a numbered box containing either a directional arrow (\texttt{SWAP} step) or a \texttt{CNOT} operation label.}
    \label{fig:routing_131}
\end{figure}

\begin{figure}[h]
    \centering
    \includegraphics[width=0.8\linewidth]{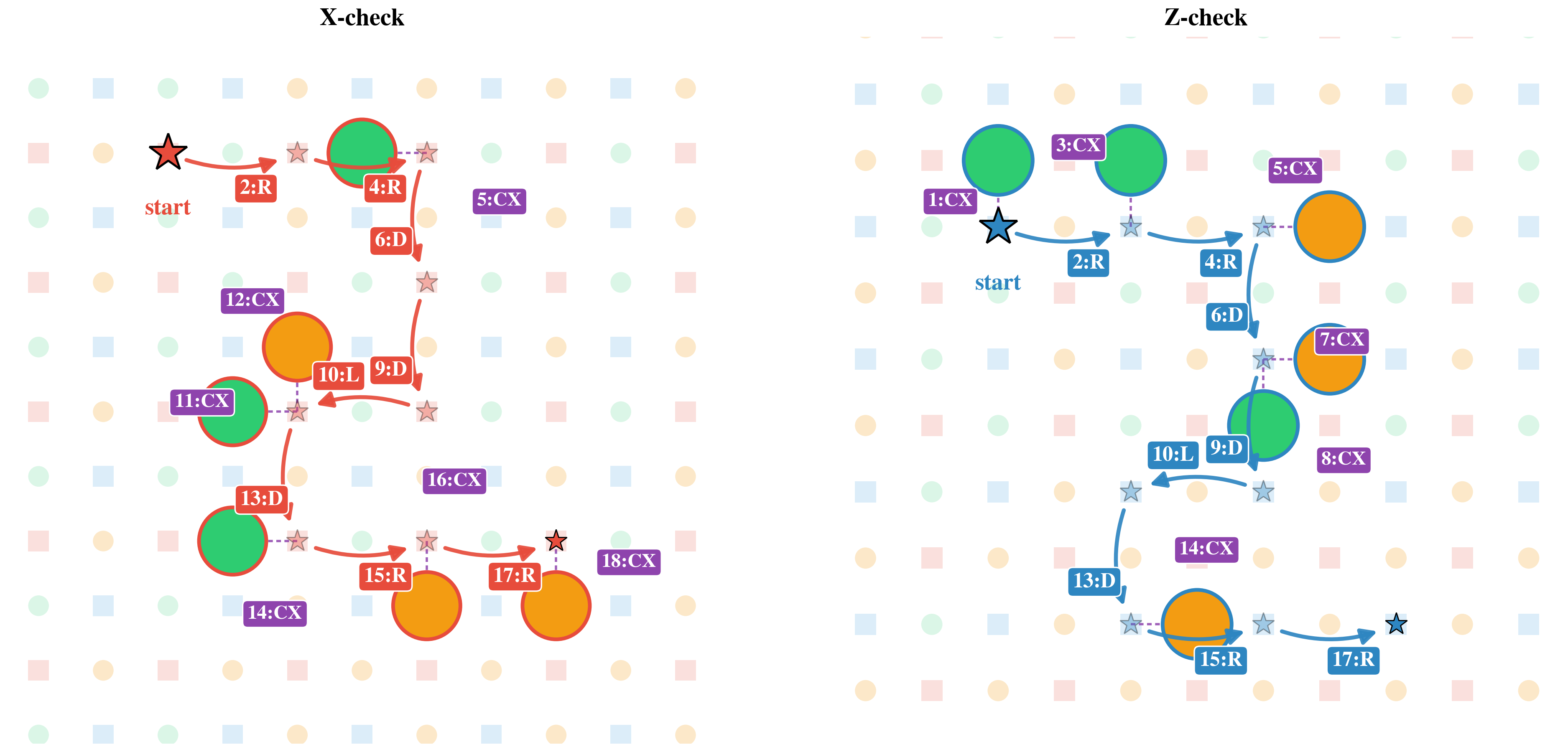}
    \caption{Routing schedule for the $[[188,8,9]]$ family. Word: \texttt{RRDDLDRR}, 18 ticks/round. Green and yellow circles indicate DataR and DataL support qubits, respectively; the ancilla is shown with a distinctive marker symbol. Each tick is represented as a numbered box containing either a directional arrow (\texttt{SWAP} step) or a \texttt{CNOT} operation label.}
    \label{fig:routing_188}
\end{figure}

In the main text, we described the algorithm for finding the routing scheme and the result found for each code family by the three-phase exhaustive search.
In this Appendix, we visualize the routing scheme in Figures~\ref{fig:routing_288}--\ref{fig:routing_188}, which show the ancilla trajectories and \texttt{CNOT} operations for one representative stabilizer from each family, in a data qubits frame.

\clearpage
\section{Thresholds under the Standard Depolarizing Model} \label{app:threshold_standard}
As in the SI1000 noise model, whose threshold plot is reported in the main text, we estimate the threshold of four tile code families under the standard depolarizing model.

\begin{figure*}[h]\centering
\includegraphics[width=\linewidth]{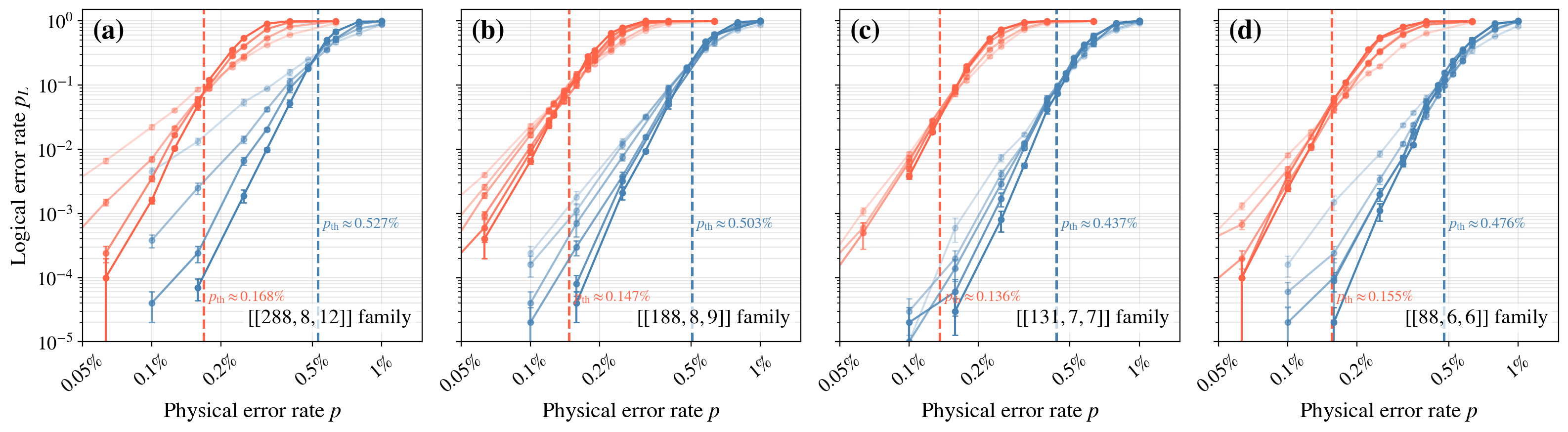}
\caption{
Logical error rate $p_L$ versus physical error rate $p$ for the (a) $[[288,8,12]]$, (b) $[[188,8,9]]$, (c) $[[131,7,7]]$, and (d) $[[88,6,6]]$ code families under the standard depolarizing noise model. Blue curves correspond to results without routing and red curves to results with \texttt{SWAP} routing. Within each panel, lighter shades indicate smaller code distances (see Table~\ref{tab:tile_code_families}). Dashed vertical lines indicate the threshold for each family obtained from finite-size scaling.
}
\label{fig:pl_standard}
\end{figure*}

\clearpage
\section{Resource Efficiency under the Standard Depolarizing Model}\label{app:footprint_standard}

\begin{figure*}[h]\centering
\includegraphics[width=0.5\linewidth]{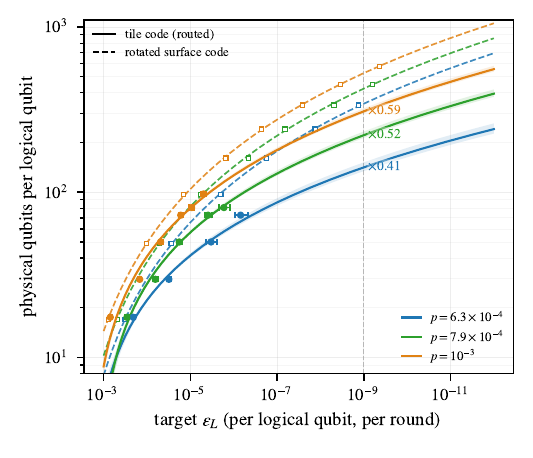}
\caption{
Physical qubits per logical qubit required to reach a target per-logical, per-round error rate $\epsilon_L$ under the standard depolarizing model, for the routed $[[288,8,12]]$ tile-code family (solid) and rotated surface-code memories (dashed). Conventions are the same as in Fig.~\ref{fig:footprint}.}
\label{fig:footprint_standard}
\end{figure*}

Figure~\ref{fig:footprint_standard} shows the footprint analysis of Sec.~\ref{sec:footprint} repeated under the standard depolarizing model, with the same per-layer idle noise acting on both codes. At $p = 10^{-3}$ the routed $[[288,8,12]]$ family has $\Lambda = 2.4$ while the surface code has $\Lambda = 7.8$, but the higher encoding rate more than makes up for this. Reaching $\epsilon_L = 10^{-6}$ takes $127$ qubits per logical qubit versus $181$ for the surface code, a $30\pm2\%$ saving. At $10^{-9}$ the tile code needs $308$ versus $525$, a $41\pm2\%$ saving, and at $10^{-12}$ the saving grows to $47\pm2\%$ ($557$ versus $1055$). The routed tile code is more qubit-efficient across the entire target range, so there is no break-even crossing at or below $p = 10^{-3}$. At the lower anchor $p = 7.9\times10^{-4}$ the advantage is larger still, the same monotonic trend as under SI1000 in the main text. The $p = 6.3\times10^{-4}$ tile curve is limited by statistics at large $d$ and is shown for reference only.

\end{document}